\documentclass{article}

    \usepackage[preprint, nonatbib]{neurips_2020}

\usepackage[utf8]{inputenc} 
\usepackage[T1]{fontenc}    
\usepackage{xr-hyper}
\usepackage{hyperref}       
\usepackage{url}            
\usepackage{booktabs}       
\usepackage{amsfonts}       
\usepackage{nicefrac}       
\usepackage{microtype}      
\usepackage{epsfig}
\usepackage{graphicx}
\usepackage{array,multirow}
\usepackage{xspace}
\usepackage{etoolbox}

\makeatletter
\DeclareRobustCommand\onedot{\futurelet\@let@token\@onedot}
\def\@onedot{\ifx\@let@token.\else.\null\fi\xspace}

\def\eg{\emph{e.g}\onedot}

\def\etal{\emph{et al}\onedot}
\makeatother

\graphicspath{{./fig/}}

\patchcmd{\maketitle}{\@fnsymbol}{\@alph}{}{}  

\title{Noise2Stack: Improving Image Restoration by Learning from Volumetric Data}

\author{%
    Mikhail Papkov\thanks{Institute of Computer Science, University of Tartu (Estonia, Tartu, Narva mnt. 18)} \\
    \texttt{mikhail.papkov@ut.ee} 
    \And Kenny Roberts\footnotemark[2] \\
    \texttt{kr19@sanger.ac.uk} 
    \And
    Lee Ann Madissoon\thanks{Wellcome Sanger Institute (United Kingdom, Hinxton, Saffron Walden CB10 1RQ)} \\
    \texttt{lee.ann.madissoon@gmail.com}
    \And Omer Bayraktar\footnotemark[2]\\
    \texttt{ob5@sanger.ac.uk}
    \And 
    Dmytro Fishman\footnotemark[1] \\
    \texttt{dmytro.fishman@ut.ee}
    \And
    Kaupo Palo\thanks{PerkinElmer Cellular Technologies (Estonia, Tallinn, Vabaõhumuuseumi tee 2a-29)}\\
    \texttt{kaupo.palo@perkinelmer.com}
    \And 
    Leopold Parts\footnotemark[1]\, \footnotemark[2]\\
    \texttt{leopold.parts@ut.ee} \\
}

\begin{document}

\maketitle

\begin{abstract}
  Biomedical images are noisy. The imaging equipment itself has physical limitations, and the consequent experimental trade-offs between signal-to-noise ratio, acquisition speed, and imaging depth exacerbate the problem. Denoising is, therefore, an essential part of any image processing pipeline, and convolutional neural networks are currently the method of choice for this task. One popular approach, Noise2Noise, does not require clean ground truth, 
  and instead, uses a second noisy copy as a training target. 
  Self-supervised methods, like Noise2Self and Noise2Void, relax data requirements by learning the signal without an explicit target, 
  but are limited by the lack of information in a single image. Here, we introduce Noise2Stack, an extension of the Noise2Noise method to image stacks that takes advantage of a shared signal between spatially neighboring planes. Our experiments on magnetic resonance brain scans and newly acquired multiplane microscopy data show that learning only from image neighbors in a stack is sufficient to outperform Noise2Noise and Noise2Void and close the gap to supervised denoising methods. 
  Our findings point towards low-cost, high-reward improvement in the denoising pipeline of multiplane biomedical images. 
  As a part of this work, we release a microscopy dataset to establish a benchmark for the multiplane image denoising.
  
\end{abstract}

\section{Introduction and related work}
\label{sec:intro}

Noise is inevitable in biological and medical imaging. Equipment imperfection in microscopy leads to image artifacts, some signal frequencies are undersampled in magnetic resonance scans, exposure time is restricted to capture dynamic information, or to avoid sample bleaching and phototoxicity. Therefore, image analysis in medicine and biology often relies on image reconstruction. The reconstruction's primary objective is to approximate "clean" image signal $s = x + s'$ by recovering absent information $s'$ from the acquired image $x$, or decomposing the acquisition $x = s + n$ into signal $s$ and degrading noise $n$. Given that the ways to approach both reconstruction tasks are similar, we will further use the terms "reconstruction", "restoration" and "denoising" interchangeably.

Deep learning advances have improved image restoration~\cite{lehtinen2018noise2noise, krull2019noise2void, krull2019probabilistic, buchholz2019cryo, batson2019noise2self, weigert2018content}. Convolutional neural networks are currently outperforming traditional (non-trainable) methods such as non-local means~\cite{buades2011non} or BM3D~\cite{dabov2007image} in open benchmarks~\cite{lemarchand2019opendenoising}, \eg~in fluorescent microscopy denoising~\cite{zhang2019poisson}. From the deep learning point of view, the most intuitive way to approach the task is to train a neural network to reconstruct clean target from the noisy image. This approach is often referred to as Noise2Clean. During the training, network parameters are optimized to minimize the error between predicted signal $\hat{s}$ and true signal $s$. But acquiring clean ground truth is challenging, often requiring multiple measurements and their subsequent registration ~\cite{weigert2018content, zhang2019poisson}. 

Perhaps surprisingly, training is also possible without clean ground truth. One way to train a denoising neural network in such way is to use two independently corrupted versions (i.e. multiple acquisitions) of the same image, as was proposed by Lehtinen~\etal~\cite{lehtinen2018noise2noise}. Noise2Noise reuses corrupted versions of the same image both as input and target to effectively reconstruct the clean signal across various noise types and image domains including magnetic resonance imaging (MRI) scans~\cite{lehtinen2018noise2noise}. This approach is mathematically justified by an assumption that the network learns the expected value of target pixels (mean, median or mode — depending on used loss function). One of its main limitations is the requirement to have two independently corrupted copies of the data.
While it is relatively straightforward to synthetically corrupt ground truth images twice, acquiring two independent measurements is usually not feasible.

To mitigate this requirement, multiple methods were developed to train a denoising network from a single corrupted measurement. Noise2Void by Krull~\etal~\cite{krull2019noise2void} and Noise2Self by Batson and Royer~\cite{batson2019noise2self} exploited the idea that signal is not statistically independent across the image, but noise is, so self-supervision could be applied. Noise2Void is trained to predict a central pixel under the blind-spot from a patch around it, while Noise2Self generalizes the approach by introducing an additional masking procedure. Laine~\etal~\cite{laine2019improved} further improved blind-spot networks and performed on par with Noise2Noise. 
Noisier2Noise proposed by Moran~\etal~\cite{moran2019noisier2noise} outperformed Noise2Void in self-supervised denoising but required a predefined model of the noise. Finally, Noise2Void has been extended into a probabilistic model by Krull~\etal~\cite{krull2019probabilistic} by estimating the parameters of the noise distribution from either camera characteristics measurement or noise-free copy of the data, which allowed to compete with supervised models on a publicly available microscopy dataset~\cite{zhang2019poisson}. 

\begin{figure}[h]
\begin{center}
\includegraphics[width=0.45\linewidth]{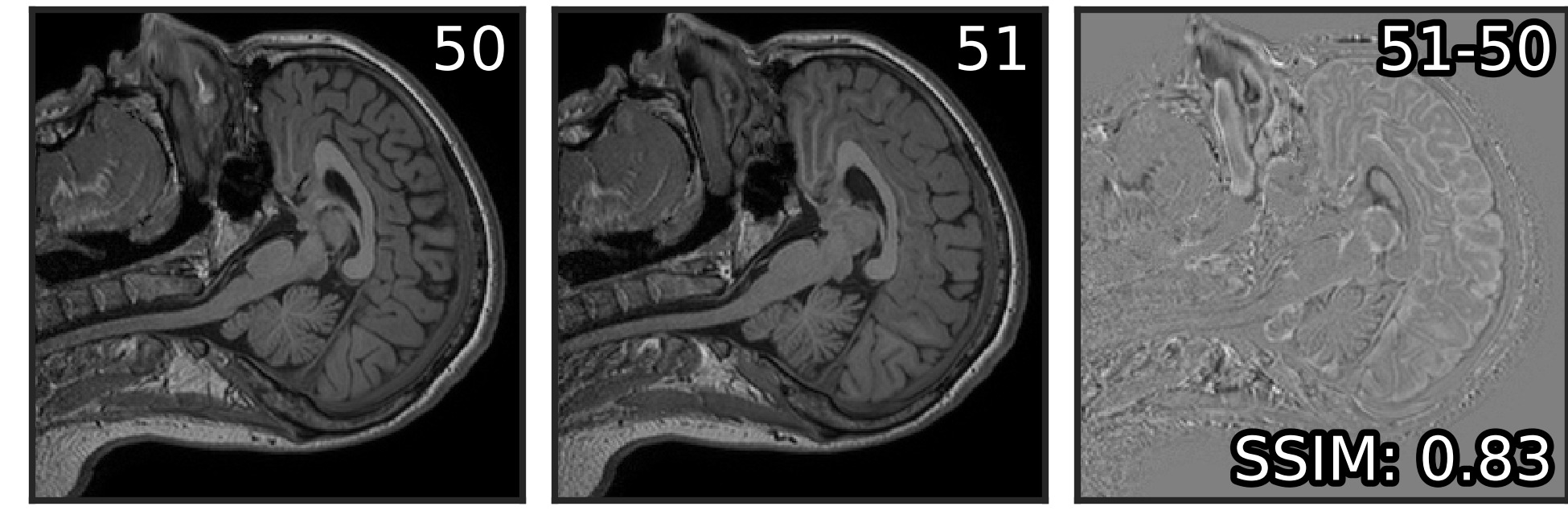}
\hspace{10pt}
\includegraphics[width=0.45\linewidth]{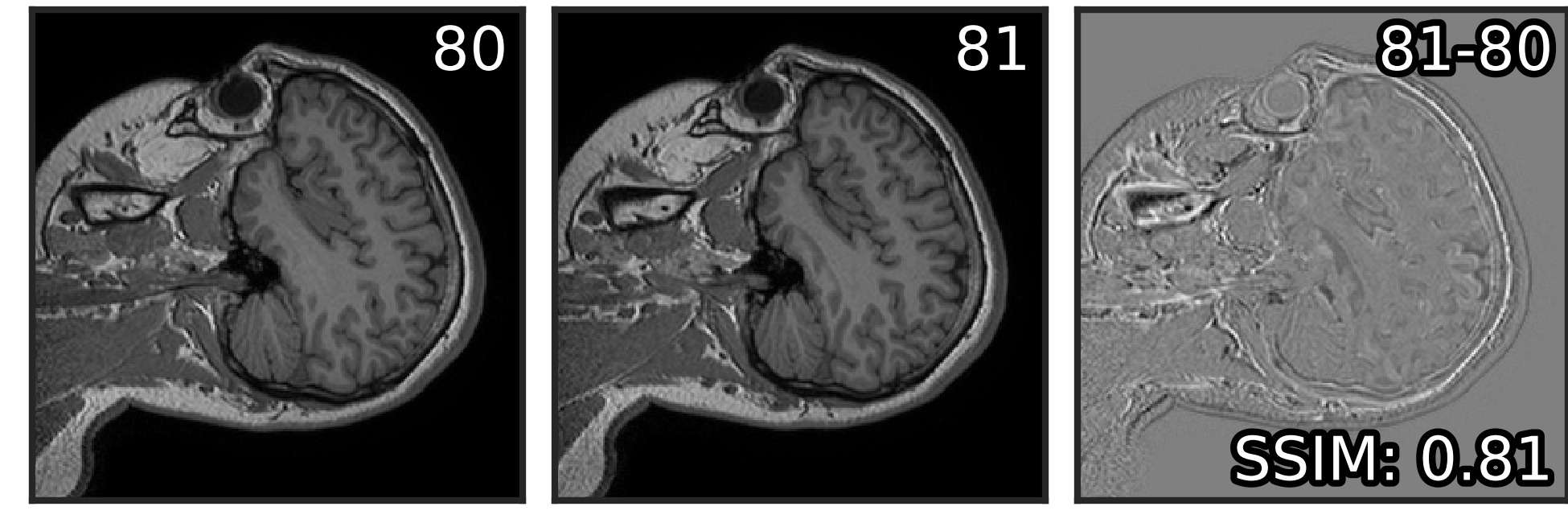}

\end{center}
   \caption{Structural similarity (SSIM) of neighboring planes in MRI dataset. Planes 50 and 51 from one stack, followed by the residual between neighboring planes; and the same for planes 80 and 81. All images are scaled in range $[-0.5, 0.5]$. Residuals have near-zero means and standard deviations 0.05 and 0.07 respectively. SSIM between planes 50 and 80 is $0.38$. High structural similarity suggests that we can use plane neighbors for reconstruction, but within limited distance.}
\label{fig:ssim-show}
\end{figure}

There are opportunities to improve these models by using recent discoveries about neural network architectures, loss functions and training procedures. However, while extended approaches described above dominated traditional training-free denoising algorithms, they could not outperform supervised methods like Noise2Noise which use more information. This motivates investigating ways to use more data during training instead. In health imaging and cell biology, it is common to have a stack of 2D planes that could be combined in a single volumetric image. For MRI, these planes are resonance signals from a certain sample depth~\cite{lustig2008compressed}, for cell microscopy, acquisitions from consecutive focal planes~\cite{dalgarno2010multiplane}. Given that each image consists of multiple planes, it may also be possible to take advantage of mutual information between neighboring planes and denoise an image stack from self (see Figure~\ref{fig:ssim-show}). This approach was applied to denoise cryogenic electron microscopy (cryo-EM)~\cite{buchholz2019cryo}, low-dose computer tomography (CT) and MRI data~\cite{wu2019consensus}. There, a stack of tomograms was split into even and odd planes, and a network was trained to predict even planes using odd and vice versa. This approach is beneficial when neither clean data nor the second noisy copy is available. 
On the other hand, the number of training examples in this case decreases to the number of available image stacks, which often is a strict limiting factor in neural network training.

In this work, we present Noise2Stack, an extension of the Noise2Noise model that makes use of information from the neighbouring planes to improve the quality of the image denoising (see Figures~\ref{fig:title},~\ref{fig:microscopy} for example results). 
Our experiments demonstrate that Noise2Stack trained on neighbouring planes in addition to the copy of the target image, consistently outperforms the original Noise2Noise method. Moreover, we show that even without noisy copy of the target image, its performance matches or even exceeds the one of Noise2Noise method. Such training scheme allows learning from a single stack of 2D images without requiring the second independently corrupted copy, which is often hard or impossible to acquire.


\section{Methods}

\subsection{Noise2Stack}
\label{sec:n2s}

Our first main contribution is the Noise2Stack method. It is a sampling algorithm for training denoising neural networks which selects a set of planes from a provided stack on each step. For example, in order to denoise the $i$-th plane in the stack, the network can be given information from the neighbors $i+1$ and $i-1$. Below we describe the nuances of our algorithm's work.

Noise2Stack can operate in two modes. In the  copy-supervised mode, a copy of the reconstructed plane is passed as an input along with its neighbors. Without adding neighbouring planes, this is exactly equivalent to the original Noise2Noise strategy.
In the second, self-supervised mode, the reconstructed plane is held out as a training target and not used as an input. For even numbers of input planes, the self-supervised mode was used. For example, with two-plane training, to denoise the $i$-th plane, only its neighbors $i+1$ and $i-1$ were used. For odd numbers of planes, the copy-supervised mode was used: the $i$-th plane was also a part of the input. 

\begin{figure}[h]
\begin{center}
\includegraphics[width=0.85\linewidth]{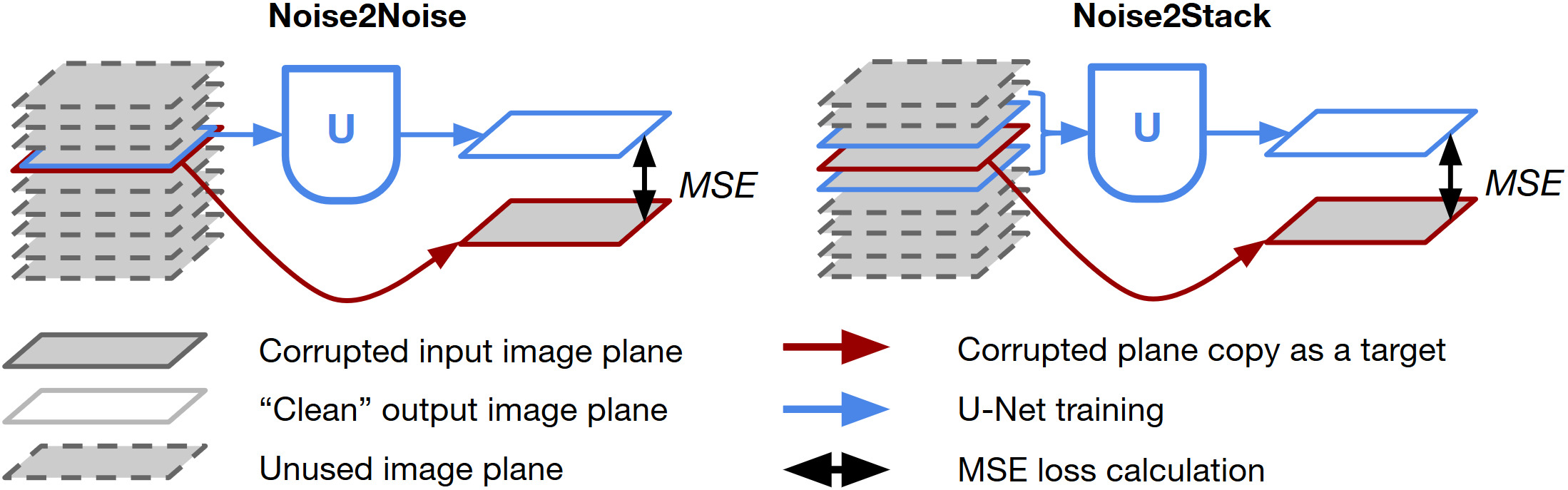}
\end{center}
   \caption{Noise2Noise and Noise2Stack training strategies. In Noise2Noise, two independently corrupted copies of each image plane are used (blue copy as an input, red copy as a target). In Noise2Stack we can choose whether we show reconstructed plane to the network as an input along with its neighbors (two neighbors training illustrated) or not. It gives an opportunity to train the network when the second corrupted copy of the stack is not available. Unused image planes are excluded from the training process at this step.}
\label{fig:model}
\end{figure}

In both modes, we train a single denoising neural network for all the planes ignoring their absolute spatial location in a stack as in Noise2Noise. Note that for the self-supervised mode, we do not require two independently corrupted copies of a stack. Instead, we use just a single noisy stack, predicting each image plane using its neighbors. For the marginal planes, we replace their absent neighbors with copies of their nearest neighbor plane (\eg~for the first plane we use two copies of the second plane). Figure~\ref{fig:model} illustrates the difference between Noise2Noise and Noise2Stack in a self-supervised mode.

\subsubsection{Neural network architecture}
\label{sec:nn}


Denoising can be done with different network architectures which often are domain-specific. 
The previously proposed Noise2Noise~\cite{lehtinen2018noise2noise} and Noise2Void~\cite{krull2019noise2void} approaches are not architectures \textit{per se}, but training strategies, which can be applied with any suitable neural network backbone. Noise2Stack is no exception to this rule. 
The only architectural aspect one may need to modify when adopting a new backbone is the number of input channels, which in general case is equal to the number of planes sampled from the stack at each training step.
In our experiments, we used previously published variations of a U-Net~\cite{ronneberger2015u} that were applied to similar tasks.
For MRI denoising, we used an architecture presented in the Noise2Noise paper~\cite{lehtinen2018noise2noise}, and for microscopy denoising we adopted an architecture from CSBDeep framework~\cite{weigert2018content}. Both of them were implemented using PyTorch deep learning library~\cite{paszke2019pytorch} in Python, and both are described in Supplementary material (Tables~\ref{tab:supp-network-microscopy},~\ref{tab:supp-network-mri}).

\subsubsection{Training}
\label{sec:training}

In all the experiments the neural network was trained for 100 epochs with batch size 16 on a single NVIDIA V100 GPU. As in~\cite{lehtinen2018noise2noise}, we did not use any regularization techniques, such as dropout or batch normalization. The network was trained using Adam~\cite{kingma2017adam} optimizer with initial learning rate $10^{-3}$ and default parameters $\beta_1=0.9$, $\beta_2=0.999$, $\epsilon = 10^{-8}$ without weight decay. The learning rate was decreasing every epoch according to the cosine schedule~\cite{loshchilov2016sgdr}. We used mean squared error (MSE) loss function as recommended in~\cite{lehtinen2018noise2noise}. The loss was computed directly on the network output without additional post-processing steps.

\subsection{Microscopy dataset}
\label{sec:dataset}

To test our approach on multiplane microscopy images, we acquired a novel dataset~\cite{kenny_roberts_2020_4114086}. CHO, U2OS, and RPE1 cell lines were grown in standard medium, seeded on plates, fixed with formaldehyde, stained with a fluorescent dye (Hoechst33342) and imaged with PerkinElmer Phenix high-throughput confocal microscope in fluorescence and brightfield modalities using a 20x objective. For the fluorescent modality, images were acquired in low-exposure (20 ms) and high-exposure (100 ms) modes. For the brightfield, we used a single exposure (100 ms). Higher exposure time usually translates into better image quality, alleviating Poisson noise, however, it leads to sample degradation, and lower measurement speed.

In total, three wells on a plate were imaged, one well per cell line, each with nine fields of view. Each field of view was imaged across ten focal planes separated by 2 micrometers in z-stack, resulting in 270 microscopy images of size $2160 \times 2160$ pixels. We used the first two wells (CHO and U2OS, 180 images) as a training set, the first three fields of view from the last well (RPE1, 30 images) as a validation set, and the last six fields of view (RPE1, 60 images) as a test set. 

\subsection{Noise generation and post-processing for MRI}
\label{sec:noise-gen}
\label{sec:post}

To generate noise for MRI dataset, we followed the procedure presented in~\cite{lehtinen2018noise2noise}. We used a Bernoulli process to sample from the frequency space of the image signal by performing a Fourier transform, and retaining each frequency $k$ with probability $p(k) = e^{-\lambda |k|}$. Signal from selected frequencies was weighted with the inverse of their respective probabilities, and omitted frequencies were replaced with zeros. As in~\cite{lehtinen2018noise2noise}, the sampling parameter $\lambda$ was chosen to preserve 10\% of frequencies relative to a full sampling. Resulting spectra were then transformed back to the original space by inverse Fourier transform.

The noise coming from spectrum undersampling provides an opportunity for post-processing.
For MRI images, we transformed the network's input and output into frequency space, and copied all the  sampled frequencies from the input image spectrum to the output as is. After copying frequencies, we transformed the resulting combined spectrum back to the image with inverse fast Fourier transform. As was shown in~\cite{lehtinen2018noise2noise}, enforcing the exact preservation of input frequencies in such a way improves the result of image restoration. This post-processing was done under the assumption that we need to restore only those frequencies which were masked. As such post-transformation is possible only when the noise comes from frequency undersampling and is not feasible for \eg~Poisson or Gaussian noise in microscopy, we report the performance both before and after post-processing for MRI.

\subsection{Metrics}
\label{sec:metrics}

We used peak signal-to-noise ratio (PSNR) as a primary metric to compare our results with previous reports~\cite{lehtinen2018noise2noise}. In addition, we measured structural similarity index (SSIM) and normalized root mean squared error (NRMSE). For fluorescent microscopy images, before measuring metrics we applied percentile normalization and affine transformation as described in~\cite{weigert2018content}. First, we percentile-normalized ground truth between $0.1$ and $99.9$, then we affinely scaled and translated predictions to minimize MSE between predictions and ground truth. Data range for metric calculation was set to 1.


\section{Experiments}
\label{sec:res}

We evaluated our Noise2Stack method against several baselines on two multiplane datasets from different biomedical domains. First, we tested the method's ability to remove the synthetic noise from MRI in the full IXI brain scan T1 dataset.\footnote{http://biomedic.doc.ic.ac.uk/brain-development/downloads/IXI/IXI-T1.tar} Noise was generated by undersampling frequencies as described in \ref{sec:noise-gen}. Second, we denoised multiplane low-exposure microscopy images in brightfield and fluorescent modalities. For the fluorescent images, we evaluated the denoising results against registered high-exposure images and in the downstream segmentation task.

\subsection{MRI}
\label{sec:res-mri}

 MRI dataset consists of volumetric images of $256 \times 256$ pixels with 150 planes each, from 60 subjects in total. We selected 100 middle planes from the full stack since marginal planes contain less information and are not as informative for noise measurements. We selected 4800 planes from 48 subjects for the training set, 200 planes from 2 subjects for the validation set and 1000 planes from 10 subjects for the test set. The list of subjects selected for testing matches the validation set from~\cite{lehtinen2018noise2noise}. Our validation set is a small subset of the training set from~\cite{lehtinen2018noise2noise} and is used to select the best model based on the validation loss prior to testing. During pre-processing pixel intensities from each image were scaled and shifted to range $[-0.5, 0.5]$, no additional normalization was applied. As an only data augmentation technique, we randomly translated images in the range $[0, 64]$ pixels .

\label{sec:res-baseline}
We measured the baseline performance on the whole test set for direct inverse fast Fourier transform reconstruction of undersampled spectra as in~\cite{lehtinen2018noise2noise}. Our test set had an average PSNR of $20.7$ dB for directly reconstructed images. In addition, we directly applied the post-processing step by combining the spectra of input and output images. For the copy-supervised Noise2Noise training, this is possible because we assume that we have two independently corrupted copies of data. We consider this noise level more appropriate to evaluate improvement in image quality for strategies that use two image copies. When two input images spectra are combined, mean PSNR rises to $22.4$ dB (Table~\ref{tab:results}). 

\begin{figure}[h]
\begin{center}
\includegraphics[width=0.9\linewidth]{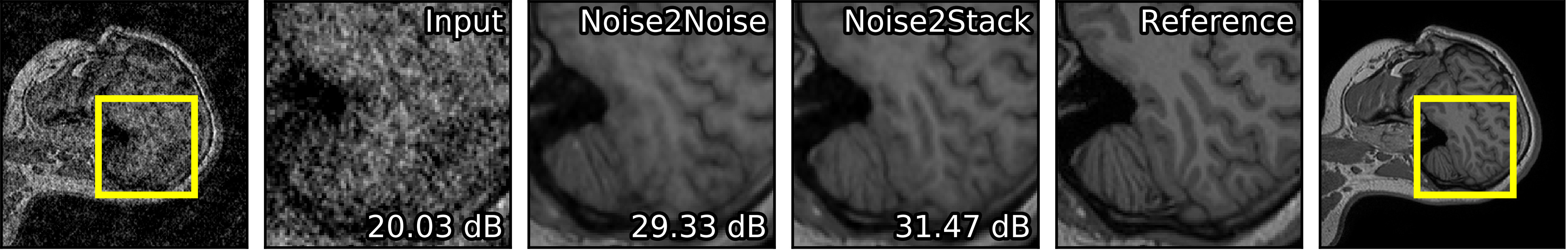}
\end{center}
   \caption{Results of Noise2Stack MRI denoising compared to Noise2Noise, input image and reference ground truth. Peak signal-to-noise ratio in bottom right corner is calculated against reference image. }
\label{fig:title}
\end{figure}

\label{sec:res-n2n}
To ensure the correctness of the following experiments and architecture implementation, we replicated the results of Noise2Noise paper~\cite{lehtinen2018noise2noise}. Here, independently corrupted copies of a single image plane were used as an input and as an output. A single plane Noise2Noise model increases PSNR to $29.1$, and to $30.9$ with further post-processing. Our test set results for the original Noise2Noise training strategy are slightly ($0.8$ dB) worse than previously reported on the validation dataset~\cite{lehtinen2018noise2noise}. 

\begin{figure}[h]
\begin{center}
\includegraphics[width=0.99\linewidth]{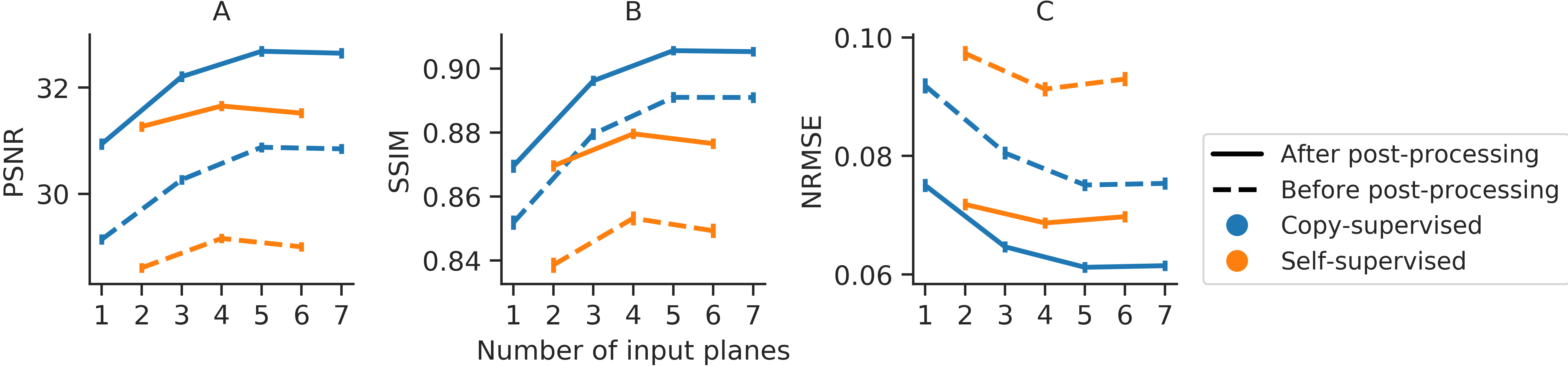}
\end{center}
  \caption{Noise suppression metrics dependent on the number of planes used in training for magnetic resonance imaging (MRI) data. Peak signal-to-noise ratio averaged across the test set images (PSNR~$\uparrow$; panel A), structural similarity (SSIM~$\uparrow$; panel B), normalized root mean squared error (NRMSE~$\downarrow$; panel C; all y-axis) for increasing number of planes used in the training (x-axis) using a copy-supervised (blue), or self-supervised (orange) approach with post-processing (solid line) and without (dashed line), training details are disclosed in text. 
   Vertical bars: 95\% confidence intervals. }
\label{fig:psnr-ssim}
\end{figure}

\label{sec:res-n2s}
Adding four neighbors to the Noise2Noise input increased PSNR by $1.8$ dB both before and after post-processing (Table~\ref{tab:results}). Overall, limiting the number of neighbouring planes to four (two from each side) proved optimal, resulting in the greatest performance benefits with reasonable computational costs in both modes (Figure~\ref{fig:psnr-ssim}). Using a single image stack for training in a self-supervised mode with four neighbours in the input, we outperform the original Noise2Noise model by $0.8$ dB.

Next, we investigated whether position in the stack affected denoising performance. We evaluated the best trained model across different plane locations and found that the marginal planes have higher PSNR after the reconstruction as well as for the baseline (Figure~\ref{fig:to-baseline}A).

\begin{table}[h]
\centering
\caption{Trained model performance on magnetic resonance imaging (MRI) and fluorescence cell microscopy datasets. For MRI, metrics were measured against clean image copy without synthetic noise. For microscopy, metrics were measured against high-exposure sample which we considered clean. For each model average results before and after post-processing are shown alongside. 
}
\label{tab:results}
\begin{tabular}{lllcccccc} \toprule
    Dataset & Denoising method  & $N$ input & \multicolumn{2}{c}{PSNR $\uparrow$}  & \multicolumn{2}{c}{SSIM $\uparrow$} & \multicolumn{2}{c}{NRMSE $\downarrow$} \\
    && planes & before & after & before & after & before & after \\ \midrule
      MRI & - & -            & 20.7   & 22.4  & 0.38   & 0.43  & 0.244  & 0.201 \\ \cmidrule{2-9}
    & Noise2Noise &     1      & 29.1   & 30.9  & 0.85   & 0.87  & 0.092  & 0.075 \\ \cmidrule{2-9}
    & Noise2Stack &    1+2            & 30.3   & 32.2  & 0.88   & 0.90  & 0.080   & 0.064 \\ 
    & (copy-supervised) &    1+4            & 30.9   & 32.7  & 0.89   & 0.91  & 0.075  & 0.061 \\
    &  &    1+6            & 30.8   & 32.6  & 0.89   & 0.91  & 0.075  & 0.061 \\ \cmidrule{2-9}
    &  Noise2Stack&    2            & 28.6   & 31.3  & 0.84   & 0.87  & 0.097  & 0.072 \\ 
    &  (self-supervised)&    4            & 29.2   & 31.7  & 0.85   & 0.88  & 0.091  & 0.069 \\
    &  &    6            & 29.0     & 31.5  & 0.85   & 0.87  & 0.093  & 0.070 \\ \midrule
    Microscopy & - & - & 25.8 && 0.40 && 0.388 & \\ \cmidrule{2-9}
     & BM3D & 1 & 31.0 && 0.68 && 0.222 & \\ \cmidrule{2-9}
     & Noise2Clean & 1 & 32.8 && 0.77 && 0.174 & \\ \cmidrule{2-9}
    & Noise2Void & 1 & 26.8 && 0.44 && 0.357 & \\ \cmidrule{2-9}
    & Noise2Stack & 2 & 31.6 && 0.73 && 0.195 & \\ 
    & (self-supervised) & 4 & 31.6 && 0.73 && 0.197 & \\ \bottomrule 
\end{tabular}

\end{table}

\subsection{Microscopy}

\begin{figure}[h]
\begin{center}
\includegraphics[width=0.4\linewidth]{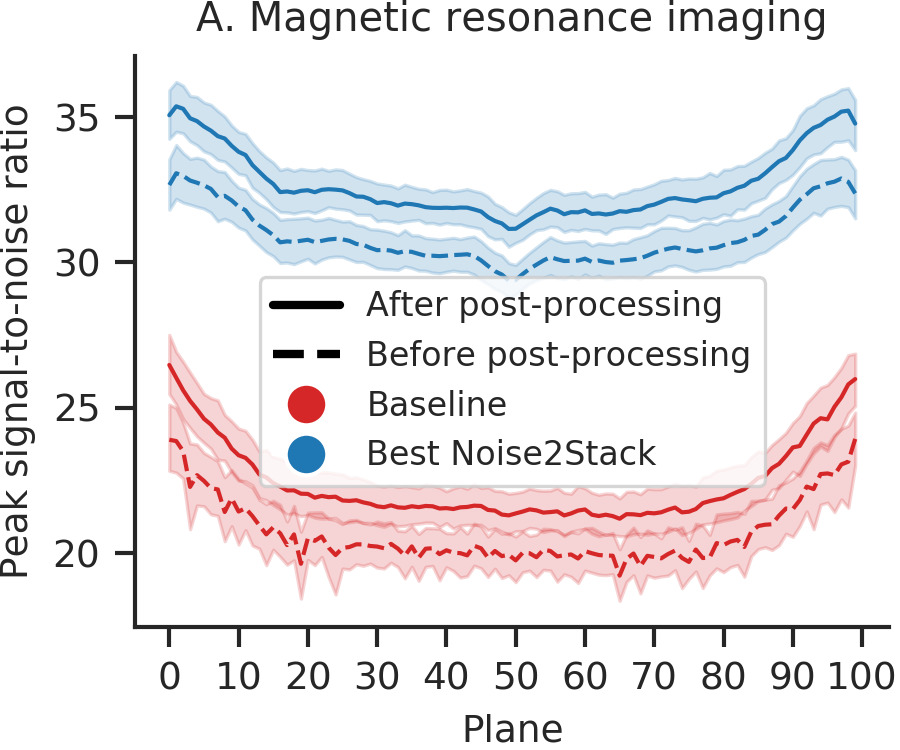}
\hspace{10pt}
\includegraphics[width=0.4\linewidth]{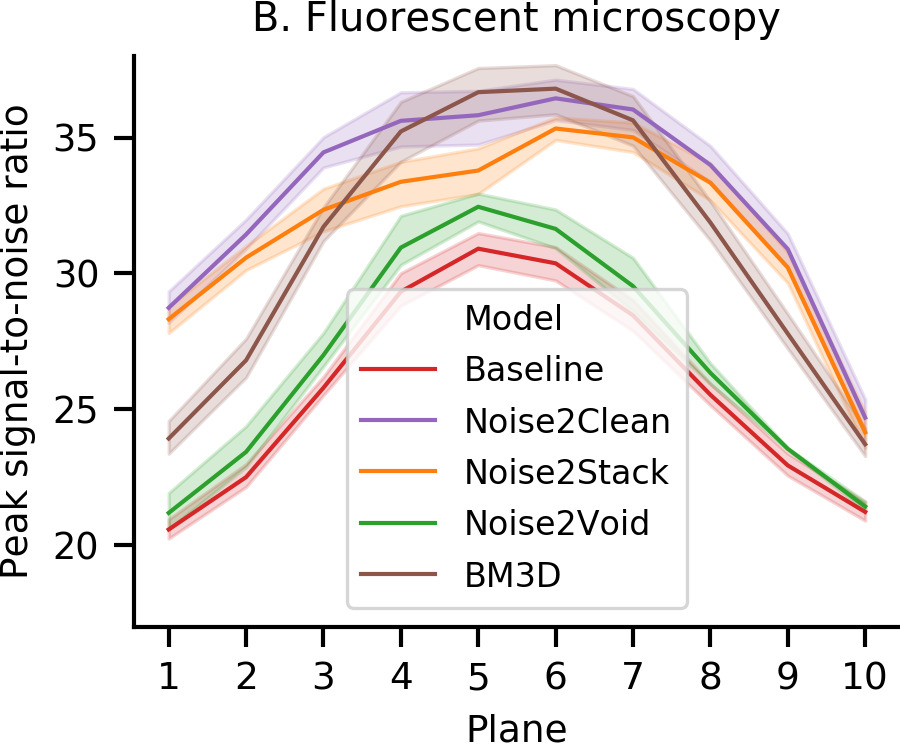}
\end{center}
   \caption{The impact of plane location on model performance. 
   \textbf{A:} Average peak signal-to-noise ratio on the test set of MRI dataset (y-axis), ranging the target plane location across the 100 available ones (x-axis). Red: baseline (direct inverse Fourier transform); blue: copy-supervised Noise2Stack with five input planes (two neighbors from each side); error band: 95\% confidence interval for 10 subjects. 
   \textbf{B:} Average peak signal-to-noise ratio on the test set of fluorescent microscopy dataset (y-axis), ranging the target plane location across the 10 available ones (x-axis). Red: baseline (low-exposure image); other colors: models in comparison; error band: 95\% confidence interval for 6 stacks.}
\label{fig:to-baseline}
\end{figure}

Microscopy dataset consists of 270 pairs of low-exposure and high-exposure fluorescent modality images and 270 respective brightfield images. We split them 180/30/60 between training/validation/test sets. The detailed dataset description provided in Section~\ref{sec:dataset}. Before the training, camera background pattern was calculated as a median across all images in each modality, and subtracted from each image before normalization between $3$ and $99.8$ percentiles as in~\cite{weigert2018content}. Background subtraction was done to remove correlated signal and prevent the network from learning identity mapping (Figure~\ref{fig:supp-background}). During training, images were randomly rotated by \{0, 90, 180, 270\} degrees, flipped both horizontally and vertically, and cropped to $256 \times 256$ pixels.

\begin{figure}[h]
\begin{center}
\includegraphics[width=0.98\linewidth]{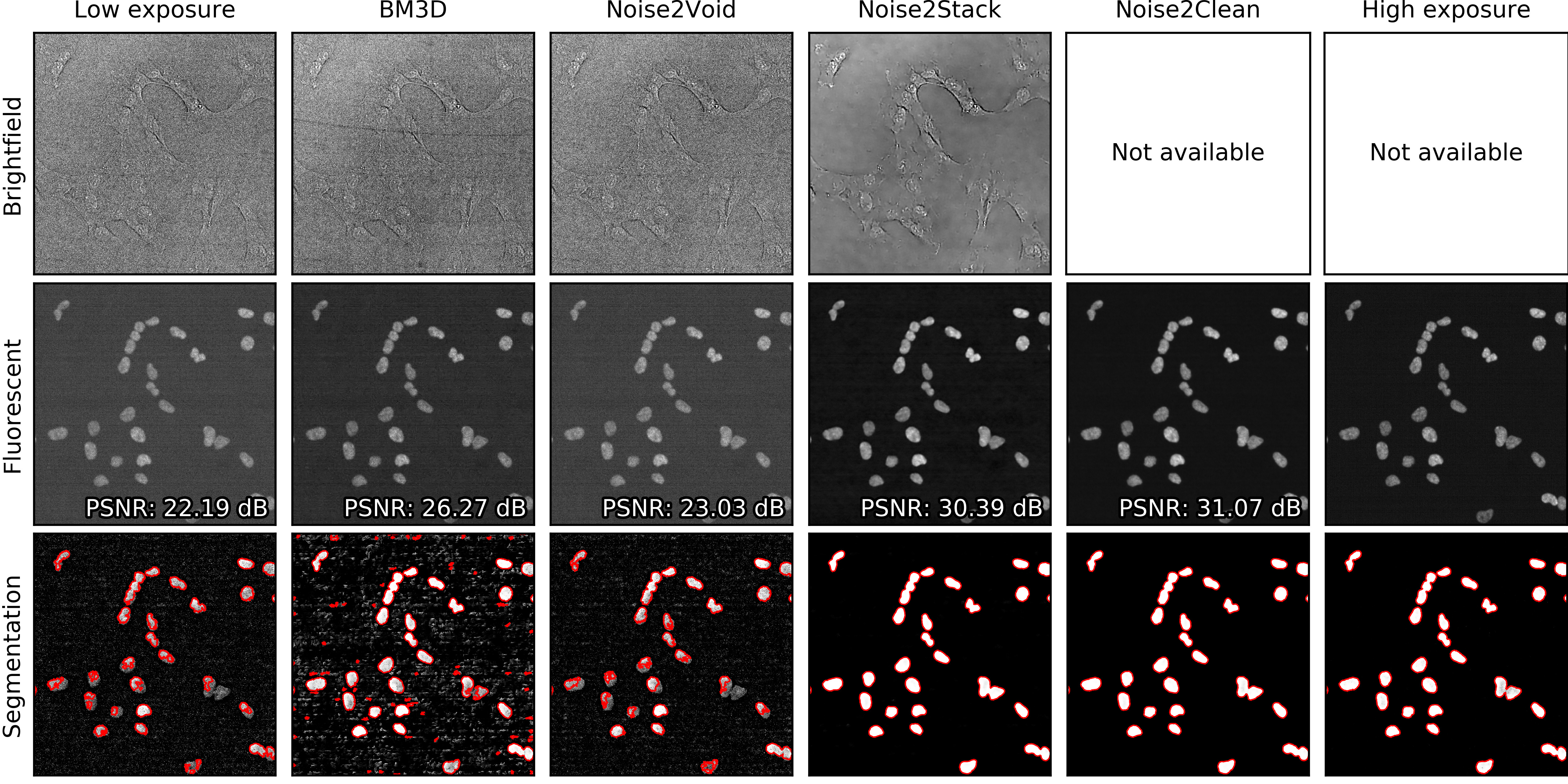}
\end{center}
   \caption{Microscopy denoising results compared to low-exposure and high-exposure images in fluorescent and brightfield modalities (the second plane from the fourth test stack was taken as an illustrative example). Segmentation of fluorescent images with a pre-trained neural network illustrates denoising effects on downstream analysis; red: probability map thresholded by $0.5$ and cleaned.}
\label{fig:microscopy}
\end{figure}

We compared Noise2Stack with supervised denoising method, Noise2Clean, which uses low-exposure images as inputs and high-exposure images as training targets. The second stack of low-exposure images was not available, therefore, training Noise2Noise network was not possible. In addition, we evaluated our strategy against another self-supervised method, Noise2Void~\cite{krull2019noise2void}, which does not require a noise model, does not put any restrictions on volumetric data properties, and treats all the planes independently of their position in the stack. For the training, we used the default Noise2Void configuration recommended by the authors in their GitHub repository. As a baseline, we used the unsupervised BM3D denoising algorithm~\cite{dabov2007image, makinen2019exact}.

High-exposure fluorescent images were used as a ground truth when denoising low-exposure images in supervised way and calculating metrics. Brightfield images naturally exhibit higher levels of noise comparing to fluorescence, thus, 100 ms exposure images could not possibly be used as clean targets. Therefore, we only evaluated brightfield denoising results visually. 

Figure~\ref{fig:to-baseline}B illustrates peak signal-to-noise ratio for different models as well as for raw images with respect to the plane position in a stack. Experiment results are summarized in Table~\ref{tab:results}. Unlike MRI data, four-plane input did not show considerable improvement comparing to two-plane mode. Thus, in both Figures~\ref{fig:to-baseline} and~\ref{fig:microscopy} we showed denoising results obtained using two neighbouring planes as input.
Figure~\ref{fig:microscopy} shows visual comparison of the denoising results with low-exposure (noisy) and high-exposure (clean) microscopy images. Noise2Void has shown only moderate improvement comparing to the original images adding $1$ dB to the baseline PSNR of $25.8$ dB. At the same time Noise2Stack was able to capture the signal in a self-supervised way, achieving PSNR $31.6$ dB and producing visually almost as good results as supervised Noise2Clean, although $1.2$ dB less by PSNR. Both Noise2Stack and Noise2Clean outperformed the BM3D baseline in our experiments.

Denoising is usually done during pre-processing to ease the downstream image analysis for both machine algorithms and human researchers. To demonstrate the applicability of Noise2Stack as a general first step in the image analysis pipeline, we segmented fluorescence microscopy images with a pre-trained neural network. For pre-training we used 3072 high-exposure fluorescent images from another experiment with cell nuclei stained by the same protocol and annotated in a semi-automated way~\cite{fishman2019segmenting}. From each image we subtracted its mean value, then divided by the standard deviation. Images were segmented independently of their position in the stack one by one without any aggregation, because small nuclei could appear in lower planes and vanish in upper ones or vice versa. We did not produce manual annotation for our test data and thus did not quantify segmentation results. Instead, we visually evaluated the consistency of predictions and the amount of artifacts introduced by noise. In Figure~\ref{fig:microscopy} we show semantic segmentation probability maps for low-exposure, denoised, and high-exposure images. Segmentation of low-exposure images denoised with Noise2Clean and Noise2Stack produces plausible results, sometimes even better than segmentation of high-exposure images. We observed the biggest impact of denoising algorithms in marginal planes, where signal-to-noise ratio was extremely low. More examples from other test images in different stack positions are included in the Supplementary material (Figure~\ref{fig:supp-microscopy}).  

\section{Discussion}

The Noise2Stack method was inspired by the observation of neighboring plane similarity in a number of volumetric biomedical datasets (Figure~\ref{fig:ssim-show}). For example, for MRI brain scan dataset that we used in our experiments (Section~\ref{sec:dataset}), the average structural similarity between neighboring planes is $0.86$. We showed that U-Net is able to capture this mutual information and effectively use it for denoising. In both copy-supervised (where original Noise2Noise belongs) and self-supervised modes, two neighbors from each side (four total) were optimal for denoising of MRI. Two neighbors, one from each side, were best for the fluorescent microscopy. Adding more neighbors to the stack did not affect the performance in copy-supervised mode and made it slightly worse for self-supervised mode, where only neighbors were used as the network's input. We hypothesize that this effect depends on the physical distance between imaged planes, and that the optimal number of planes differs between datasets. However, it is clear that further planes provide less useful information for the restoration and could lead to network overfitting.


We observed the marginal planes in MRI data to have higher PSNR for both baseline and denoised images (Figure~\ref{fig:to-baseline}). Exactly the opposite is true for fluorescent microscopy images: middle planes in a stack have the highest peak signal-to-noise ratio. 
And although these patterns differ across the domains, Noise2Stack has shown stable improvement of the performance comparing to methods that do not use information from the neighbouring planes.  

Unsupervised BM3D algorithm showed good results for the middle microscopy planes, however failed in the marginal ones with lower signal-to-noise ratio (Figure~\ref{fig:to-baseline}B). In complex conditions, it produced a large number of artifacts that corrupted the downstream segmentation results (Figure~\ref{fig:microscopy}). In our experiments, it was also 700 times slower than all of the deep learning methods due to the large image size and lack of GPU speed up.  

Visually, the level of detail in MRI denoised with Noise2Stack in copy-supervised mode has noticeably increased comparing to Noise2Noise, as can be seen in Figure~\ref{fig:title} (additional illustrations provided in Supplementary material, Figure~\ref{fig:supp-showcase}). Brain gyri in the reconstructed images are clean and sharp even for the cerebellum where they are tiny. In self-supervised mode Noise2Stack results are expectedly less detailed and match Noise2Noise by sight. However, this mode demands twice less data, being able to learn from a single copy. For the fluorescent microscopy, Noise2Stack in self-supervised mode did not outperform supervised Noise2Clean, but has shown comparable results.


\section{Conclusion}
\label{sec:conclusion}

We introduced Noise2Stack — an approach to improve Noise2Noise results for volumetric data. We use neighboring image planes as an additional source of information for image restoration, and demonstrate that this boosts performance. If the target plane is not included, denoising is a result of the middle plane prediction. Further, we show that by using only neighboring planes it is possible to achieve Noise2Noise level of performance with a single image stack. 
As the network is trained to predict the expected value of each pixel, 
the method can be used even if there is no corrupted stack copy available, reducing imaging times, or avoiding additional assumptions about the noise characteristics.
Noise2Stack can be applied in all domains where measurements are made in a spatially structured manner. The limitations of sampling to secure the required amount of shared signal between acquisitions is a subject of further studies.

Our second major contribution is a dataset of matched low exposure time (noisy) and high exposure time (clean) fluorescence cell image stacks from different cell lines accompanied by the brightfield images. These freely available data, together with the results presented here, can serve as the benchmark for future methods development in the field.

\begin{ack}

This work was funded by PerkinElmer Cellular Technologies (VLTAT19682), Wellcome Trust (206194), Estonian Research Council (IUT34-4), and Estonian Centre of Excellence in IT (EXCITE) (TK148). We thank High Performance Computing Center of the Institute of Computer Science at the University of Tartu for the provided computing power.

\end{ack}

{\small
\bibliographystyle{ieee_fullname}
\bibliography{ref}
}

\clearpage

\setcounter{figure}{0}
\setcounter{table}{0}
\renewcommand{\thetable}{S\arabic{table}}
\renewcommand{\thefigure}{S\arabic{figure}}
\renewcommand{\theHtable}{Supplement.\thetable}
\renewcommand{\theHfigure}{Supplement.\thetable}

\section*{Supplementary material}

\begin{table}[h]
\centering
\caption{Network architecture for microscopy denoising~\cite{weigert2018content}. $N_{out}$ denotes the number of convolutional filters trained in the layer and therefore the number of output feature maps. $W_{out}$ denotes the width and height of output feature maps for each layer in training. Number of input channels $n$ depends on the experiment. All convolutional layers are followed by ReLU activation function except the last one which is followed by linear activation function. For concatenation and addition, ID of concatenated layer is denoted in brackets. Residual layer 19 adds the middle plane from the input to the output.} 
\begin{tabular}{lllcc}\toprule
U-Net level & ID & Layer                     & $N_{out}$& $W_{out}$ \\ \toprule 
            & 1  & Input                     & $n$      & $256$   \\ \midrule
Down 1      & 2  & Convolution $3 \times 3$   & $32$     & $256$   \\
            & 3  & Convolution $3 \times 3$   & $32$     & $256$   \\
            & 4  & Maxpool $2 \times 2$      & $32$     & $128$   \\ \midrule
Down 2      & 5  & Convolution $3 \times 3$   & $64$     & $128$   \\ 
            & 6  & Convolution $3 \times 3$   & $64$     & $128$   \\ 
            & 7  & Maxpool $2 \times 2$      & $64$     & $64$    \\ \midrule
Middle      & 8  & Convolution $3 \times 3$   & $128$    & $64$     \\ 
            & 9  & Convolution $3 \times 3$   & $64$     & $64$     \\ \midrule
Up 2        & 10 & Upsample $2 \times 2$     & $64$     & $128$   \\
            & 11 & Concatenate (6)           & $128$    & $128$   \\
            & 12 & Convolution $3 \times 3$   & $64$     & $128$   \\
            & 13 & Convolution $3 \times 3$   & $32$     & $128$  \\ \midrule
Up 1        & 14 & Upsample $2 \times 2$     & $32$     & $256$   \\
            & 15 & Concatenate (3)           & $64$     & $256$   \\ 
            & 16 & Convolution $3 \times 3$   & $32$     & $256$   \\
            & 17 & Convolution $3 \times 3$   & $32$     & $256$   \\
            & 18 & Convolution $3 \times 3$  & $1$      & $256$  \\ \midrule
Residual    & 19 & Add (middle 1)            & $1$      & $256$ \\ \bottomrule
\end{tabular}

\label{tab:supp-network-microscopy}
\end{table}

\begin{table}[h]
\centering
\caption{Network architecture for MRI denoising~\cite{lehtinen2018noise2noise}. $N_{out}$ denotes the number of convolutional filters trained in the layer and therefore the number of output feature maps. $W_{out}$ denotes the width and height of output feature maps for each layer. Number of input channels $n$ depends on the experiment. All convolutional layers are followed by LeakyReLU activation function with $\alpha = 0.1$ except the last one which is followed by linear activation function. For concatenation, ID of concatenated layer is denoted in brackets.} 
\begin{tabular}{lllcc}\toprule
U-Net level & ID & Layer            & $N_{out}$  & $W_{out}$ \\ \toprule 
            & 1  & Input            & $n$      & $256$   \\ \midrule
Down 1      & 2  & Convolution $3 \times 3$   & $48$     & $256$   \\
            & 3  & Convolution $3 \times 3$   & $48$     & $256$   \\
            & 4  & Maxpool $2 \times 2$      & $48$     & $128$   \\ \midrule
Down 2      & 5  & Convolution $3 \times 3$   & $48$     & $128$   \\ 
            & 6  & Maxpool $2 \times 2$      & $48$     & $64$    \\ \midrule
Down 3      & 7  & Convolution $3 \times 3$   & $48$     & $64$    \\
            & 8  & Maxpool $2 \times 2$      & $48$     & $32$    \\ \midrule
Down 4      & 9  & Convolution $3 \times 3$   & $48$     & $32$    \\
            & 10 & Maxpool $2 \times 2$      & $48$     & $16$    \\ \midrule
Down 5      & 11 & Convolution $3 \times 3$   & $48$     & $16$    \\
            & 12 & Maxpool $2 \times 2$      & $48$     & $8$     \\ \midrule
Middle      & 13 & Convolution $3 \times 3$   & $48$     & $8$     \\ \midrule
Up 5        & 14 & Upsample $2 \times 2$     & $48$     & $16$    \\
            & 15 & Concatenate (10) & $96$     & $16$    \\
            & 16 & Convolution $3 \times 3$   & $96$     & $16$    \\
            & 17 & Convolution $3 \times 3$   & $96$     & $16$    \\ \midrule
Up 4        & 18 & Upsample $2 \times 2$     & $96$     & $32$    \\
            & 19 & Concatenate (8)  & $144$    & $32$    \\
            & 20 & Convolution $3 \times 3$   & $96$     & $32$    \\ 
            & 21 & Convolution $3 \times 3$   & $96$     & $32$    \\ \midrule
Up 3        & 22 & Upsample $2 \times 2$     & $96$     & $64$    \\
            & 23 & Concatenate (6)  & $144$    & $64$    \\
            & 24 & Convolution $3 \times 3$   & $96$     & $64$    \\
            & 25 & Convolution $3 \times 3$   & $96$     & $64$    \\ \midrule
Up 2        & 26 & Upsample $2 \times 2$     & $96$     & $128$   \\
            & 27 & Concatenate (4)  & $144$    & $128$   \\
            & 28 & Convolution $3 \times 3$   & $96$     & $128$   \\
            & 29 & Convolution $3 \times 3$   & $96$     & $128$  \\ \midrule
Up 1        & 30 & Upsample $2 \times 2$     & $96$     & $256$   \\
            & 31 & Concatenate (1)  & $96 + n$ & $256$   \\ 
            & 32 & Convolution $3 \times 3$   & $64$     & $256$   \\
            & 33 & Convolution $3 \times 3$   & $32$     & $256$   \\
            & 34 & Convolution $3 \times 3$  & $1$      & $256$  \\ \bottomrule
\end{tabular}

\label{tab:supp-network-mri}
\end{table}


\begin{figure}[h]
\begin{center}
\includegraphics[width=0.6\linewidth]{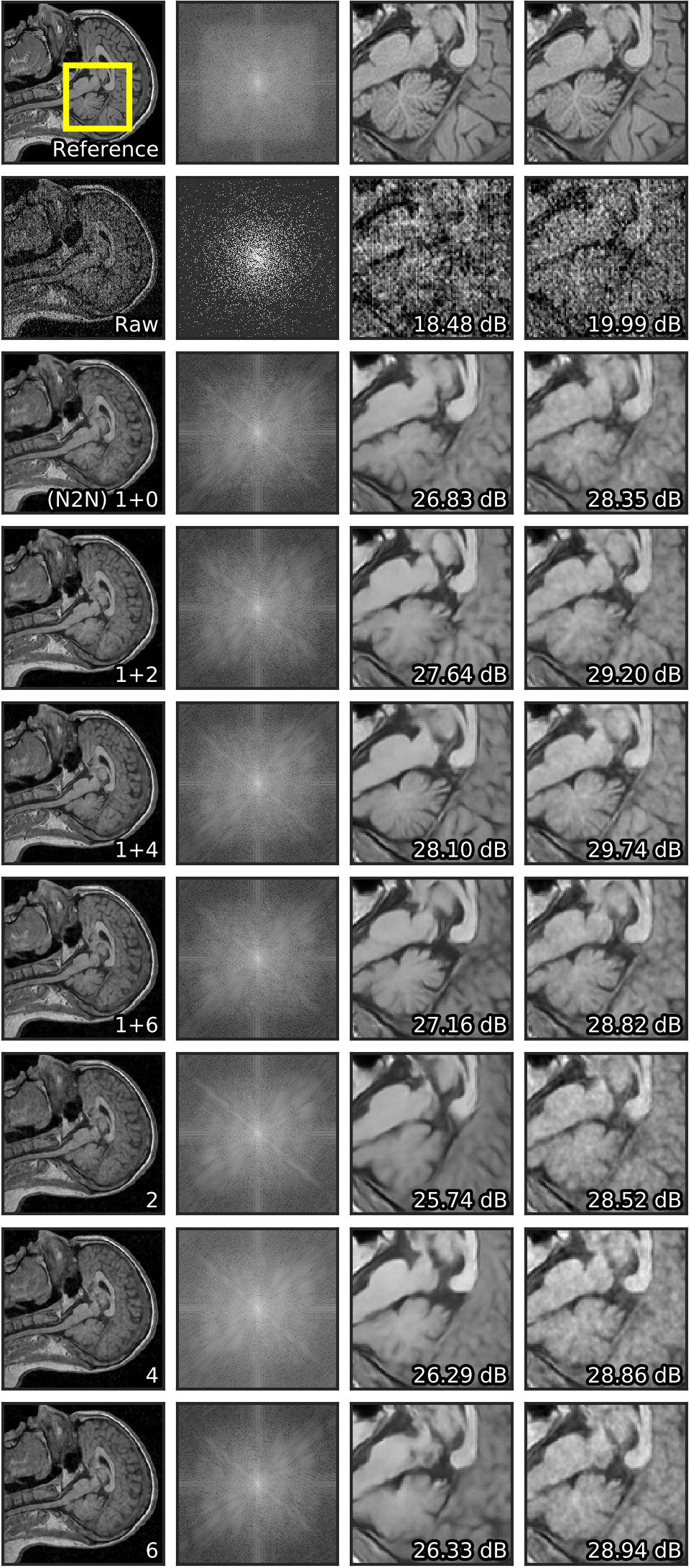}
\begin{tabular}{p{0.12\linewidth}p{0.12\linewidth}p{0.12\linewidth}p{0.12\linewidth}}
    \small \centering a)~Image  & \small \centering b)~Spectrum  & \small \centering c)~Crop  & \small \centering d)~Crop~(PP) 
\end{tabular}
\end{center}
    \caption{Example of denoising results with a different number of planes used in an input stack. Full-size image in column (a), its spectrum in (b), magnified crop from within yellow box before post-processing in (c), after post-processing in (d). PSNR is shown in the bottom right corner of the last two columns for the example image. The first row represents reference image, the second row shows raw baseline image. All the subsequent rows show the output of different models. Number of input planes for the respective model is shown in the bottom right corner in column (a). $1+N$ models were trained in a copy-supervised mode, others in self-supervised mode, $N$ denotes number of neighbors in an input stack.}
\label{fig:supp-showcase}
\end{figure}


\begin{figure}[h]
\begin{center}
\includegraphics[width=0.98\linewidth]{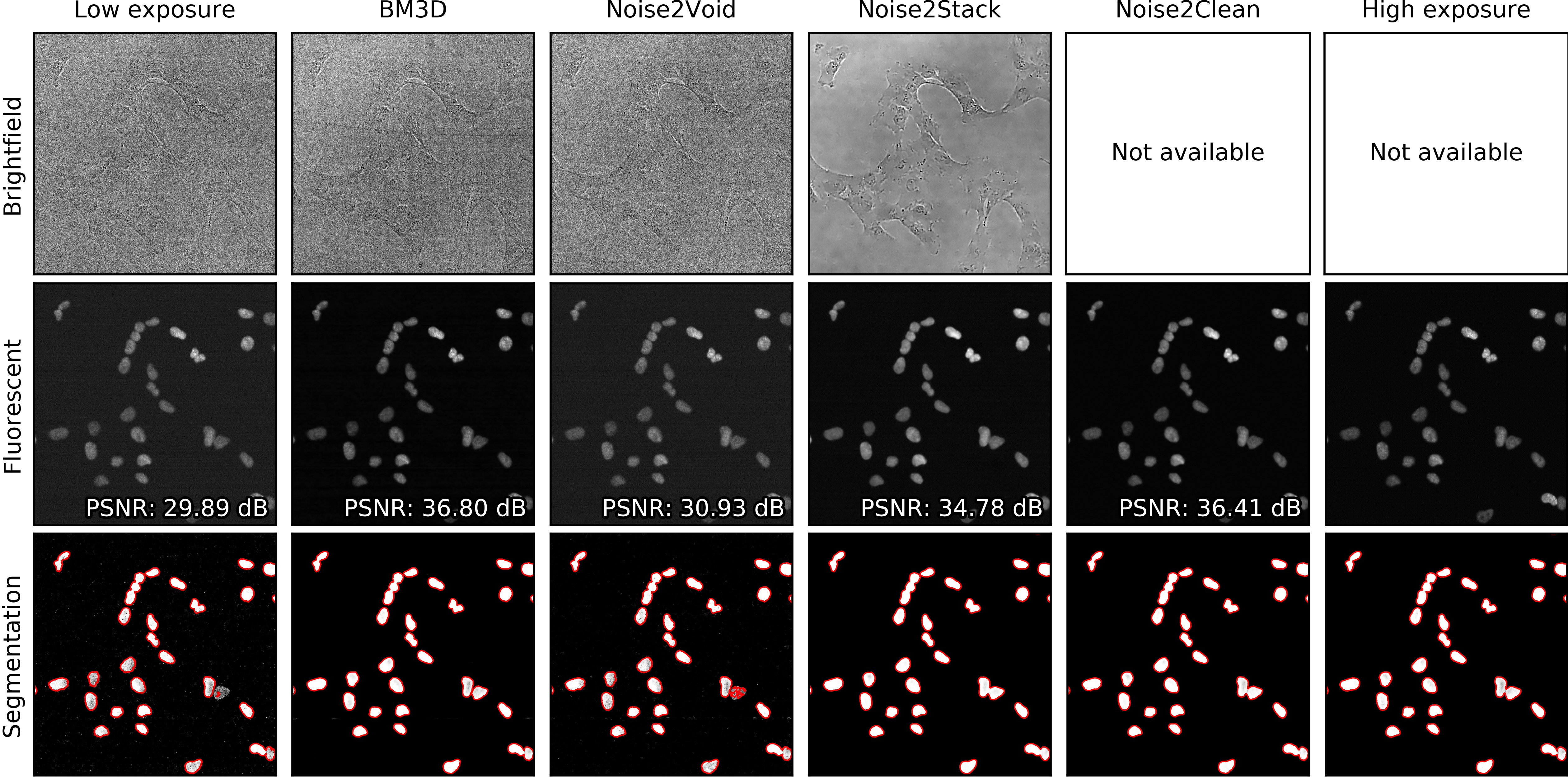}
\includegraphics[width=0.98\linewidth]{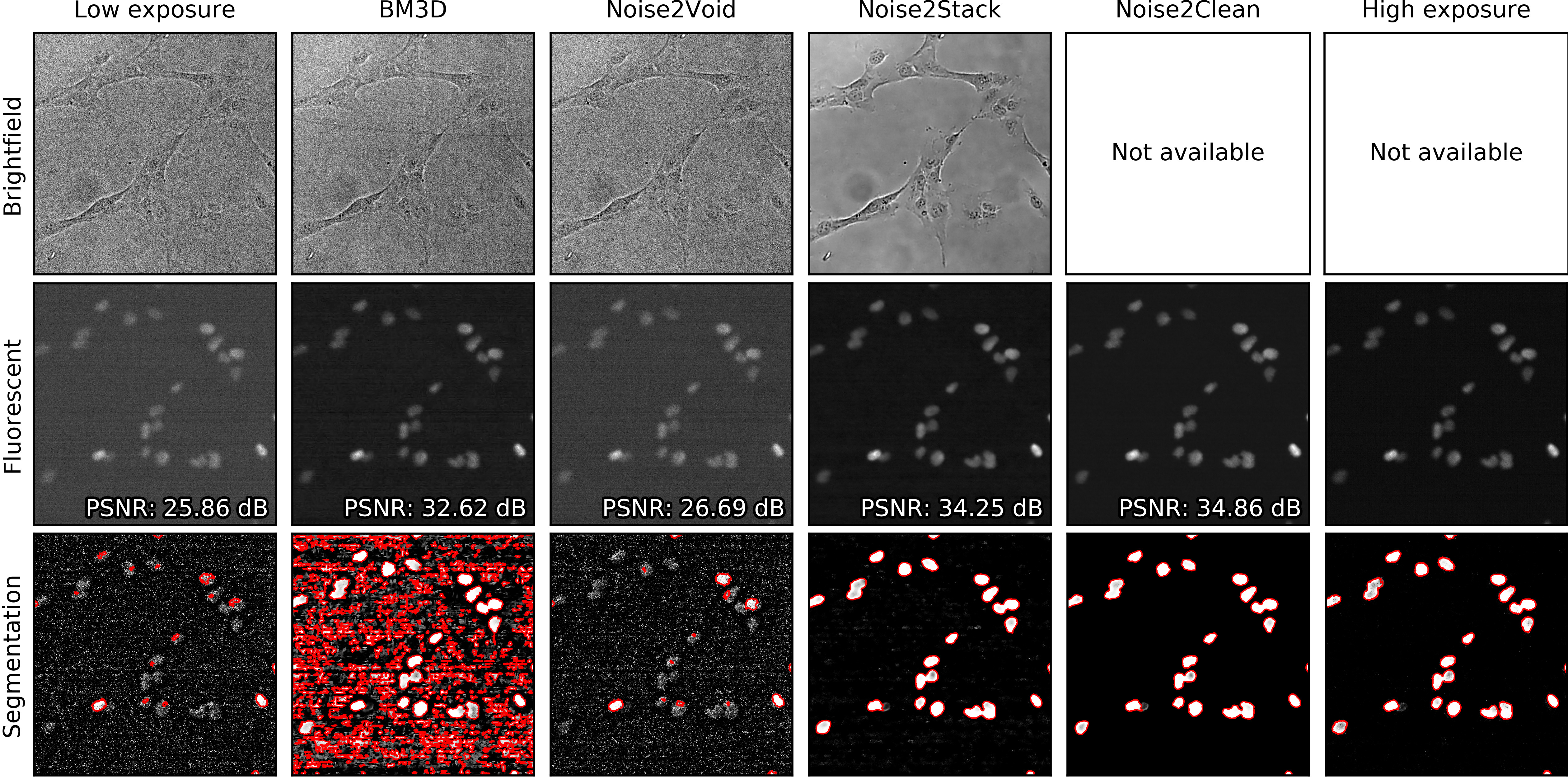}
\end{center}
   \caption{Microscopy denoising results compared to low-exposure and high-exposure images in fluorescent and brightfield modalities (top: the fifth plane from the fourth test stack; bottom: the seventh plane from the sixth test stack). Segmentation of fluorescent images with a pre-trained neural network illustrates denoising effects on downstream analysis; red: probability map thresholded by $0.5$ and cleaned.}
\label{fig:supp-microscopy}
\end{figure}

\begin{figure}[h]
\begin{center}
\includegraphics[width=0.9\linewidth]{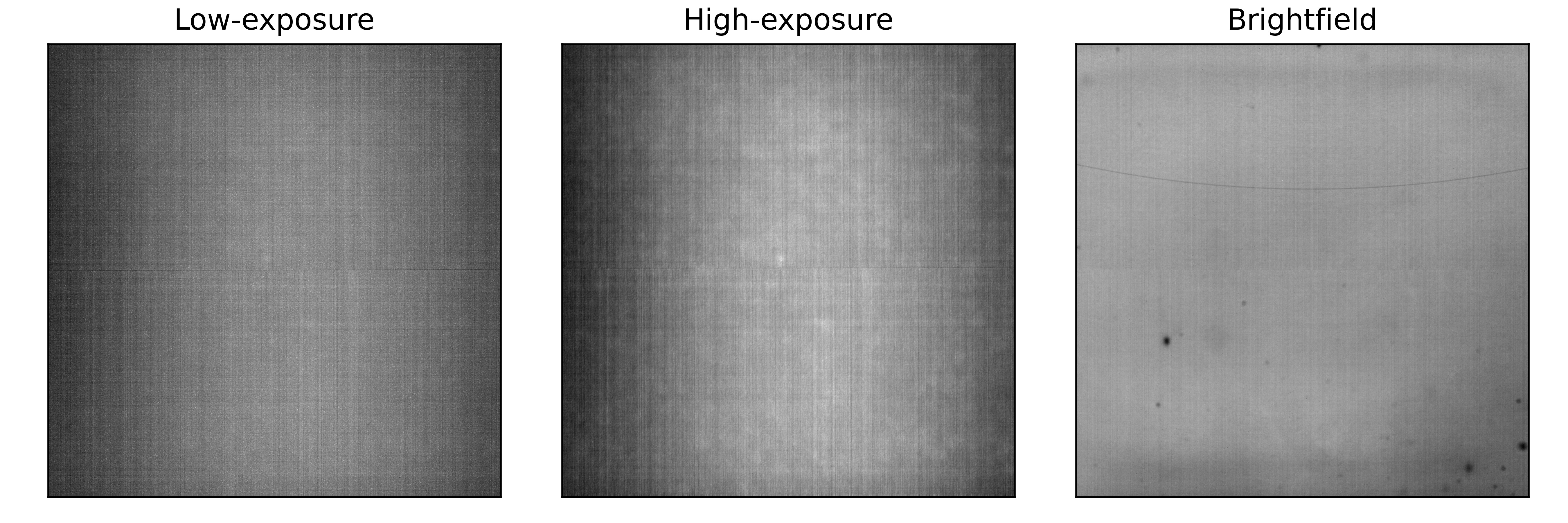}
\end{center}
   \caption{Camera background patterns in microscopy dataset calculated as a median across all low-exposure and high-exposure fluorescent, and brightfield images. Similar results could be achieved by imaging an empty well with the same microscope parameters.}
\label{fig:supp-background}
\end{figure}

\clearpage

\end{document}